\documentclass[12pt]{article}
\usepackage{amsmath}
\usepackage{amsfonts}
\usepackage{amssymb}
\usepackage{latexsym}
\usepackage{graphicx}
\usepackage[english]{babel}
\input epsf.sty

\begin{document}

\title{
{Dark Energy Model with Spinor Matter and Its Quintom Scenario}}
\vspace{3mm}
\author{{Yi-Fu Cai$^1$\footnote{caiyf@ihep.ac.cn}, Jing
Wang$^{1,2}$}\\
{\small $^{1}$ Institute of High Energy Physics, Chinese
Academy of Sciences, }\\
{\small P.O. Box 918-4, Beijing 100049, P. R. China}\\
{\small $^{2}$ Department of Physics Science, Hebei Normal
University, P. R. China} }
\date{}
\maketitle

\begin{abstract}

A class of dynamical dark energy models, dubbed Spinor Quintom,
can be constructed by a spinor field $\psi$ with a nontraditional
potential. We find that, if choosing suitable potential, this
model is able to allow the equation-of-state to cross the
cosmological constant boundary without introducing any ghost
fields. In a further investigation, we show that this model is
able to mimic a perfect fluid of Chaplygin gas with $p=-c/\rho$
during the evolution, and also realizes the Quintom scenario with
its equation-of-state across $-1$.

\end{abstract}

\maketitle

\section{Introduction}

The recent data from type Ia supernovae and cosmic microwave
background (CMB) radiation and so on \cite{Perlmutter:1996ds,
Riess:2004nr, Seljak:2004xh} have provided strong evidences for a
spatially flat and accelerated expanding universe at the present
time. In the context of Friedmann-Robertson-Walker (FRW)
cosmology, this acceleration is attributed to the domination of a
component with negative pressure, called dark energy. So far, The
nature of dark energy remains a mystery. Theoretically, the
simplest candidate for such a component is a small positive
cosmological constant, but it suffers the difficulties associated
with the fine tuning and the coincidence problems. So many
physicists are attracted by the idea of dynamical dark energy
models, such as Quintessence \cite{Ratra:1987rm}, Phantom
\cite{Caldwell:1999ew}, K-essence \cite{ArmendarizPicon:2000dh},
Quintom \cite{Feng:2004ad} and so on (see Refs.
\cite{Copeland:2006wr, Padmanabhan:2007xy} for a review). Usually,
dark energy models are constructed by scalar fields which are able
to accommodate a rich variety of behaviors phenomenologically.
However, there is another possibility that the acceleration of the
universe is driven by a classical homogeneous spinor field. Some
earlier studies on applications of spinor fields in cosmology are
given in Refs. \cite{taub:1937, Brill:1957fx, Parker:1971pt}. In
recent years there are many works on studying spinor fields as
gravitational sources in cosmology, for example: see Refs.
\cite{Obukhov:1993fd, ArmendarizPicon:2003qk} for inflation and
cyclic universe driven by spinor fields; see Refs.
\cite{Saha:2000nk} for spinor matter in Bianchi Type I spacetime;
see Refs. \cite{Ribas:2005vr, Chimento:2007fx} for a dark energy
model with spinor matter; and so on.

Although the recent fits to the data in combination of the 3-year
WMAP \cite{Spergel:2006hy}, the recently released 182 SNIa Gold
sample \cite{Riess:2006fw} and also other cosmological
observational data show remarkably the consistence of the
cosmological constant, it is worth noting that a class of
dynamical models with the equation-of-state (EoS) across $-1$ {\it
Quintom} is mildly favored \cite{Zhao:2006qg, Wang:2006ts}. In the
literature there have been a lot of theoretical studies of
Quintom-like models. For example, motivated from string theory,
the authors of Ref. \cite{Cai:2007gs} realized a Quintom scenario
by considering the non-perturbative effects of a generalized DBI
action. Moreover, a No-Go theorem has been proven to constrain the
model building of Quintom \cite{Xia:2007km} (see Ref.
\cite{Vikman:2004dc} for some earlier considerations), and
according to this No-Go theorem there are models which involve
higher derivative terms for a single scalar field
\cite{Li:2005fm}, models with vector field
\cite{ArmendarizPicon:2004pm}, making use of an extended theory of
gravity \cite{Cai:2005ie}, non-local string field theory
\cite{Aref'eva:2005fu}, and others (see e.g. \cite{Cai:2006dm,
Quintom_1, Xiong:2007cn}). The similar work applied in
scalar-tensor theory has also been studied in Ref.
\cite{Elizalde:2004mq, Boisseau:2000pr}.

Usually, a Quintom model involves a ghost field with its kinetic
term to be negative which leads to quantum instability. In this
paper we study the dark energy model with spinor matter.
Interestingly, we find that this type of model can realize the
Quintom scenario with its EoS across the cosmological constant
boundary $w=-1$ without introducing a ghost field. Instead, the
derivative of its potential with respect to the scalar bilinear
$\bar\psi\psi$, which is defined as the mass term, becomes
negative when the spinor field lies in the Phantom-like phase. If
this model can realize its EoS across $-1$ more than one time, the
total EoS of the universe can satisfy $w\geq -1$ during the whole
evolution which is required by the Null Energy Condition (NEC)
\cite{Qiu:2007fd}, and we expect to treat this process as a phase
transition merely existing for a short while. Moreover, due to a
perfect mathematic property of the spinor field, it is possible to
combine the Quintom scenario and the picture of Chaplygin gas with
the EoS evolving from $0$ to $-1$ smoothly in Spinor Quintom. In
the literature a dark energy model of Chaplygin gas has been
proposed to describe a transition from a universe filled with
dust-like matter to an accelerated expanding stage, and hence it
has been argued that the coincidence problem of dark energy may be
alleviated in this model \cite{Kamenshchik:2001cp}.

Interestingly, in our model we are able to evade the drawbacks of
considering the Phantom field, of which the kinetic energy is
negative and so is unstable in quantum level. We notice that Ref.
\cite{Kahya:2006hc} has investigated the quantum stability of a
Phantom phase of cosmic acceleration and shown that a
super-acceleration phase can be obtained by quantum effects. Our
model is different from that one since the super-acceleration is
realized by the background contribution. However, it is
interesting that both the two approaches are stable in quantum
level since the first order of perturbation theory can be defined.

This paper is organized as follows. In Section II, we simply
review the basic algebra of a spinor field in FRW universe which
is minimally coupled to Einstein's gravity. In Section III, we
present the condition for the spinor field to realize a Quintom
scenario and give some detailed examples. In Section IV, we
provide a unified model of Spinor Quintom and a perfect fluid of
Chaplygin gas by taking certain potentials. By solving the model
numerically, we will study the evolution of its EoS and fraction
of energy density. Section V is the conclusion and discussions of
our paper.

\section{Algebra of A Spinor Field}

To begin with, we simply review the dynamics of a spinor field
which is minimally coupled to Einstein's gravity (see Refs.
\cite{Weinberg,BirrellDavies,GSW} for detailed introduction).
Following the general covariance principle, a connection between
the metric $g_{\mu\nu}$ and the vierbein is given by
\begin{equation}
g_{\mu\nu}e_{a}^{\mu}e_{b}^{\nu}=\eta_{ab}~,
\end{equation}
where $e_{a}^{\mu}$ denotes the vierbein, $g_{\mu\nu}$ is the
space-time metric, and $\eta_{a b}$ is the Minkowski metric with
$\eta_{ab}={\rm diag}(1,-1,-1,-1)$. Note that the Latin indices
represents the local inertial frame and the Greek indices
represents the space-time frame.

We choose the Dirac-Pauli representation as
\begin{eqnarray}
\gamma^0= \left(\begin{array}{cccc}
1 &   0  \\
0 &   -1
\end{array}\right),~~~
\gamma^{i}=\left(\begin{array}{cccc}
0 &          \sigma_{i} \\
-\sigma_{i}&   0
\end{array}\right),
\end{eqnarray}
where $\sigma_{i}$ is Pauli matrices. One can see that the
$4\times4$ $\gamma^{a}$ satisfy the Clifford
algebra$\{\gamma^{a},\gamma^{b}\}=2\eta_{ab}$. The $\gamma^{a}$
and $e_{a}^{\mu}$ provide the definition of a new set of Gamma
matrices
\begin{equation}
\Gamma^{\mu}=e_{a}^{\mu}\gamma^{a}~,
\end{equation}
which satisfy the algebra
$\{\Gamma^{\mu},\Gamma^{\nu}\}=2g_{\mu\nu}$. The generators of the
Spinor representation of the Lorentz group can be written as
$\Sigma^{ab}=\frac{1}{4}[\gamma^{a},\gamma^{b}]$. So the covariant
derivative are given by
\begin{eqnarray}
D_{\mu}\psi&=&(\partial_{\mu}+\Omega_{\mu})\psi~,\\
D_{\mu}\bar\psi&=&\partial_{\mu}\bar\psi-\bar\psi\Omega_{\mu}~,
\end{eqnarray}
where the Dirac adjoint $\bar\psi$ is defined as $\psi^+\gamma^0$.
The $4\times4$ matrix $\Omega_{\mu}=\frac{1}{2}\omega_{\mu
ab}\Sigma^{ab}$ is the spin connection, where $\omega_{\mu
ab}=e_{a}^{\nu}\nabla_{\mu}e_{\nu b}$ are Ricci spin coefficients.

By the aid of the above algebra we can write down the following
Dirac action in a curved space-time
background\cite{ArmendarizPicon:2003qk, Ribas:2005vr,
Vakili:2005ya}
\begin{eqnarray}\label{action}
S_{\psi}&=&\int d^4 x~e~[\frac{i}{2}(\bar\psi\Gamma^{\mu}D_{\mu}
\psi-D_{\mu}\bar\psi\Gamma^{\mu}\psi)-V]~.
\end{eqnarray}
Here, $e$ is the determinant of the vierbein $e_{\mu}^{a}$ and $V$
stands for the potential of the spinor field $\psi$ and its
adjoint $\bar\psi$. Due to the requirement of covariance, the
potential $V$ only depends on the scalar bilinear $\bar\psi\psi$
and ``pseudo-scalar" term $\bar\psi\gamma^5\psi$. For simplicity
we drop the latter term and assume that there is
$V=V(\bar\psi\psi)$.

Varying the action with respect to the vierbein $e_{a}^{\mu}$, we
obtain the energy-momentum-tensor,
\begin{eqnarray}\label{EMT}
T_{\mu\nu}&=&\frac{e_{\mu a}}{e}\frac{\delta S_\psi}{\delta
e_{a}^{\nu}} \nonumber\\
&=& \frac{i}{4}[\bar\psi\Gamma_{\nu}D_{\mu}\psi+\bar\psi
\Gamma_{\mu}D_{\nu}\psi-D_{\mu}\bar\psi\Gamma_{\nu}\psi
-D_{\nu}\bar\psi\Gamma_{\mu}\psi] -g_{\mu\nu}{\cal L}_{\psi}~.
\end{eqnarray}
On the other hand, varying the action with respect to the field
$\bar\psi$, $\psi$ respectively yields the equation of motion of
the spinor,
\begin{eqnarray}
i\Gamma^{\mu}D_{\mu}\psi-\frac{\partial V}{\partial\bar\psi}=0~,~~
iD_{\mu}\bar\psi\Gamma^{\mu}+\frac{\partial V}{\partial\psi}=0~.
\end{eqnarray}

Note that, we use units $8\pi G=\hbar=c=1$ and all parameters are
normalized by $M_p=1/\sqrt{8 \pi G}$ in the letter.

\section{A universe driven by Spinor Quintom}

\subsection{Dynamics of a spinor field}

In this paper we deal with the homogeneous and isotropic FRW
metric,
\begin{equation}
ds^{2}=dt^{2}-a^{2}(t)d\vec{x}^2~,
\end{equation}
where $a$ stands for the scale factor and we choose today's scale
factor $a_0=1$. Correspondingly, the vierbein are given by
\begin{equation}
e_{0}^{\mu}=\delta_{0}^{\mu}~,~~e_{i}^{\mu}=\frac{1}{a}\delta_{i}^{\mu}~.
\end{equation}
Assuming the spinor field is space-independent, the equation of
motion reads
\begin{eqnarray}
\label{EoMa}\dot{\psi}+\frac{3}{2}H\psi+i\gamma^{0} V' \psi&=&0~,\\
\label{EoMb}\dot{\bar\psi}+\frac{3}{2}H\bar\psi-i\gamma^{0}V'
\bar\psi&=&0~,
\end{eqnarray}
where a dot denotes a time derivative `$\frac{d}{dt}$' and a prime
denotes a derivative with respect to $\bar\psi\psi$, and $H$ is
the Hubble parameter. Taking a further derivative, we can obtain:
\begin{equation}\label{solution}
\bar\psi\psi=\frac{N}{a^{3}}~,
\end{equation}
where $N$ is a positive time-independent constant and we define it
as today's value of $\bar\psi\psi$.

From the expression of the energy-momentum tensor in Eq.
(\ref{EMT}), we get the energy density and the pressure of the
spinor field:
\begin{eqnarray}
\label{density}\rho_{\psi}&=&T_{0}^{0}=V~,\\
\label{pressure}p_{\psi}&=&-T_{i}^{i}=V'\bar\psi\psi-V~,
\end{eqnarray}
where Eqs. (\ref{EoMa}) and (\ref{EoMb}) have been applied. The
EoS of the spinor field, defined as the ratio of its pressure to
energy density, is given by
\begin{equation}\label{eos}
w_{\psi}\equiv\frac{p_{\psi}}{\rho_{\psi}}=-1+\frac{V'\bar\psi\psi}{V}~.
\end{equation}

Simply taking the potential to be power-law-like $V=V_0
(\frac{\bar\psi\psi}{N})^{\alpha}$ with $\alpha$ as a nonzero
constant, we obtain a constant EoS:
\begin{eqnarray}
w_{\psi}=-1+\alpha~.
\end{eqnarray}
In this case, the spinor matter behaves like a
linear-barotropic-like perfect fluid. For example: if
$\alpha=\frac{4}{3}$, we can get $\rho_{\psi}\sim a^{-4}$ and
$w_{\psi}=\frac{1}{3}$, which is the same as radiation; if
$\alpha=1$, then $\rho_{\psi}\sim a^{-3}$ and $w_{\psi}=0$, this
component behaves like normal matter.

Furthermore, the spinor matter is able to realize the acceleration
of the universe if $\alpha<\frac{2}{3}$. So it provides us a
potential motivation to construct a dynamical dark energy model
with the spinor matter. Moreover, as introduced at the beginning
of Section I, there is evidence in the recent observations to
mildly support a Quintom scenario with the EoS of dark energy
across $-1$. In the following, we emphasize our investigation on
constructing Quintom dark energy model with the spinor field,
which is called Spinor Quintom.

\subsection{Evolutions of Spinor Quintom}

To keep the energy density positive, one may see that there is
$w_{\psi}>-1$ when $V'>0$ and $w_{\psi}<-1$ when $V'<0$ from Eq.
(\ref{eos}). The former corresponds to a Quintessence-like phase
and the latter stands for a Phantom-like phase. Therefore it
requires the derivative of the potential $V'$ to change its sign
if one expects a Quintom picture. In terms of the variations of
$V'$, it shows
that there exists three categories of evolutions in Spinor Quintom:\\
(i), there is
\begin{eqnarray}
V'>0~~~\rightarrow~~~V'<0 \nonumber~,
\end{eqnarray}
which gives a Quintom-A scenario by describing the universe
evolving from Quintessence-like phase with $w_{\psi} > -1$ to
Phantom-like phase with $w_{\psi} < -1$; \\
(ii), there is
\begin{eqnarray}
V'<0~~~\rightarrow~~~V'>0 \nonumber~,
\end{eqnarray}
which gives a Quintom-B scenario for which the EoS is arranged to
change from below $-1$ to above $-1$; \\
(iii), $V'$ changes its sign for more than one time, then one can
obtain a new Quintom scenario with its EoS crossing $-1$ many
times, dubbed Quintom-C scenario.

In the following, we will take different potentials of Spinor
Quintom to provide the three kinds of evolutions mentioned above
\footnote{Note that we choose the potentials phenomenologically
without any constraints from quantum field theory or other
consensus. From the phenomenological viewpoint, this is okay if we
treat the background classically while deal with the perturbations
in quantum level, just as what is done in inflation theory.}.

To begin with, we shall investigate Case (i) and provide a
Quintom-A model. We use the form of potential
$V=V_{0}[(\bar\psi\psi-b)^{2}+c]$, where $V_{0}$, b, c are
undefined parameter. Then we get $V' = 2V_0(\bar\psi\psi-b)$ and
the EoS:
\begin{eqnarray}
w_{\psi} = \frac { (\bar\psi\psi)^{2} - b^{2} - c} {
(\bar\psi\psi)^{2} - 2b\bar\psi\psi + b^{2} + c} ~.
\end{eqnarray}

\begin{figure}[htbp]
\begin{center}
\includegraphics[scale=0.9]{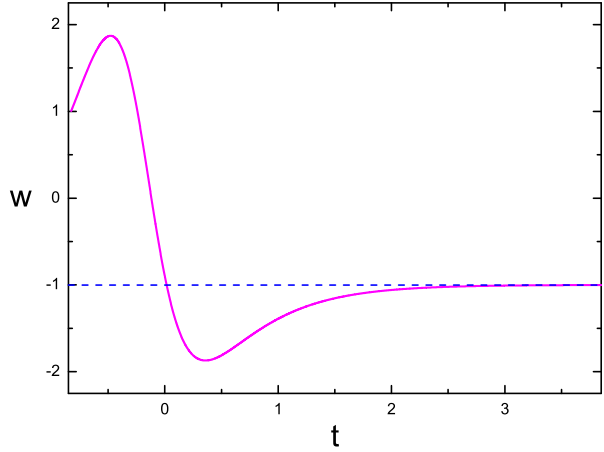}
\caption{Plot of the evolution of the EoS in Case (i) as a
function of time. In the numerical calculation we take the
potential of the spinor field as $V =
V_{0}[(\bar\psi\psi-b)^{2}+c]$, where we choose
$V_0=1.0909\times10^{-117}$, $b=0.05$ and $c=10^{-3}$ for the
model parameters. For the initial condition we take $N=0.051$.
\label{Fig:Q-A}}
\end{center}
\end{figure}

According to Eq. (\ref{solution}), one finds that $\bar\psi\psi$
is decreasing along with an increasing scale factor $a$ during the
expansion of the universe. From the formula of $V'$, we deduce
that at the beginning of the evolution the scale factor $a$ is
very small, so $\bar\psi\psi$ becomes very large and ensures
$V'>0$ at the beginning. Then $\bar\psi\psi$ decreases along with
the expanding of $a$. At the moment of $\bar\psi\psi=b$, one can
see that $V'=0$ which results in the EoS $w_{\psi}=-1$. After that
$V'$ becomes less than $0$, the universe enters a Phantom-like
phase. Finally the universe approaches a de-Sitter spacetime. This
behavior is also obtained by the numerical calculation and shown
in Fig. \ref{Fig:Q-A}. From this figure, one can read that the EoS
$w_{\psi}$ starts the evolution from $1$, then mildly increases to
a maximum and then begin to decrease. When $\bar\psi\psi=b$, it
reaches the point $w_{\psi}=-1$ and crosses $-1$ from above to
below smoothly. After that, the EoS sequentially decreases to a
minimal value then increases and eventually approaches the
cosmological constant boundary.

\begin{figure}[htbp]
\begin{center}
\includegraphics[scale=0.9]{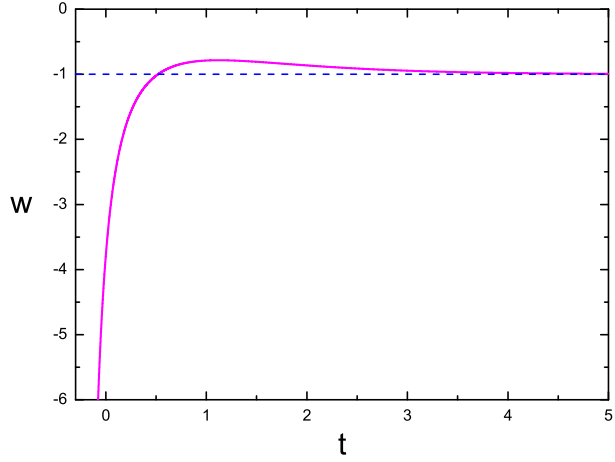}
\caption{Plot of the evolution of the EoS in Case (ii) as a
function of time. In the numerical calculation we take the
potential of the spinor field as $V = V_0
[-(\bar\psi\psi-b)\bar\psi\psi+c]$, where we choose
$V_0=1.0909\times10^{-117}$, $b=0.05$ and $c=10^{-3}$ for the
model parameters. For the initial condition we take $N=0.051$.
\label{Fig:Q-B}}
\end{center}
\end{figure}

In Case (ii) we choose the potential as $V = V_0
[-(\bar\psi\psi-b)\bar\psi\psi+c]$. Then one can obtain
$V'=V_0(-2\bar\psi\psi+b)$ and the EoS
\begin{eqnarray}
w_{\psi} =
\frac{-(\bar\psi\psi)^2-c}{-(\bar\psi\psi)^2+b\bar\psi\psi+c}~.
\end{eqnarray}
Initially $V'$ is negative due to the large values of
$\bar\psi\psi$. Then it increases to $0$ when
$\bar\psi\psi=\frac{b}{2}$ whereafter changes its sign and becomes
larger than $0$, in correspondence with the Case (ii), dubbed
Quintom-B model. From Fig. \ref{Fig:Q-B}, we can see that the EoS
evolves from below $-1$ to above $-1$ then finally approaches to
$-1$.

\begin{figure}[htbp]
\begin{center}
\includegraphics[scale=0.9]{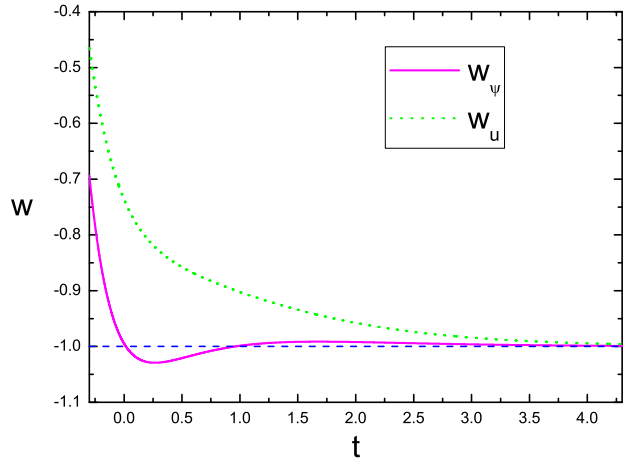}
\caption{Plot of the evolution of the EoS in Case (iii) as a
function of time. The magenta solid line stands for the EoS of
Spinor Quintom and the green dot line stands for that of the whole
universe. In the numerical calculation we take the potential of
the spinor field as $V=V_{0}[(\bar\psi\psi-b)^{2}\bar\psi\psi+c]$,
where we choose $V_0=1.0909\times10^{-117}$, $b=0.05$ and
$c=10^{-3}$ for the model parameters. For the initial condition we
take $N=0.051$. \label{Fig:quinc}}
\end{center}
\end{figure}

In the third case we explore a Quintom scenario which gives the
EoS across $-1$ for two times in virtue of the potential
\begin{eqnarray}
V=V_{0}[(\bar\psi\psi-b)^{2}\bar\psi\psi+c]~.
\end{eqnarray}
From the expression of the potential, one have
\begin{eqnarray}
V'=V_0(\bar\psi\psi-b)(3\bar\psi\psi-b)~,
\end{eqnarray}
and the EoS
\begin{eqnarray}
w_{\psi}=\frac{2(\bar\psi\psi)^3-2b(\bar\psi\psi)^2-c}
{(\bar\psi\psi)^3-2b(\bar\psi\psi)^2+b^2(\bar\psi\psi)+c}~.
\end{eqnarray}
Evidently the equation $V'=0$ has two solutions which are
$\bar\psi\psi=b$ and $\bar\psi\psi=\frac{b}{3}$, thus $V'$ changes
its sign two times. From the expression of the EoS, we find that
$w_{\psi}>-1$ in the beginning. When the value of $\bar\psi\psi$
equals to b, it crosses $-1$ for the first time. After the first
crossing, it enters the phantom-like state and continuously
descends until passes through its minimum, then ascends to
$\bar\psi\psi=\frac{b}{3}$ and then experiences the second
crossing, and eventually moving up to the Quintenssence-like
phase. This is shown in Fig. \ref{Fig:quinc}. One can see that the
big rip can be avoided in this case.

Moreover, taking the component of normal matter with the energy
density $\rho_{m} \propto 1/a^3$ into consideration and the EoS
$w_m=0$, we can see that the EoS of the universe
$w_{u}=w_{\psi}\frac{\rho_{\psi}}{\rho_{\psi}+\rho_{m}}$ satisfies
the relation $w_u\geq-1$ in this case. As is argued in Ref.
\cite{Qiu:2007fd}, NEC might be satisfied in the models though
$w_{\psi} < -1$ only stays for a short period during the evolution
of the universe.

One important issue of a dark energy model is the analysis of its
perturbations. To study this issue, we might be able to learn to
what degree the system is stable both in quantum and classical
level. Usually systems with $w<-1$ show some nasty instabilities,
for example, a Phantom universe suffers a Big rip singularity.
Although this singularity can be avoided in most Quintom models
which usually enter a de-Sitter expansion in the final epoch, all
the scalar models of Quintom by now suffer a quantum instability
since there are negative kinetic modes from ghost fields. Here we
would like to show the perturbation theory of Spinor Quintom
crudely. Since we do not introduce any ghost fields in our model,
the crossing of phantom divide is achieved by the spinor field
itself and does not perform any particular instabilities.

In order to simplify the derivative, we would like to redefine the
spinor as $\psi_N\equiv a^{\frac{3}{2}}\psi$. Then perturbing the
spinor field, one gives the perturbation equation as follows,
\begin{eqnarray}\label{perteq}
\frac{d^2}{d\tau^2}\delta\psi_N-\nabla^2\delta\psi_N+a^2\left[
V'^2+i\gamma^0 (HV'-3HV''\bar\psi\psi)
\right]\delta\psi_N\nonumber\\
=-2a^2V'V''\delta(\bar\psi\psi)\psi_N-i\gamma^\mu\partial_\mu[a
V''\delta(\bar\psi\psi)]\psi_N~,
\end{eqnarray}
where $\tau$ is the conformal time defined by $d\tau\equiv dt/a$.
Since the right hand side of the equation decays proportional to
$a^{-3}$ or even faster, we can neglect those terms during the
late time evolution of the universe for simplicity.

From the perturbation equation above, we can read that the sound
speed is equal to $1$ which eliminates the instability of the
system in short wavelength. Moreover, when the EoS $w$ crosses
$-1$, we have $V'=0$ at that moment and the eigen function of the
solution to Eq. (\ref{perteq}) in momentum space is a Hankel
function with an index $\frac{1}{2}$. Therefore, the perturbations
of the spinor field oscillate inside the hubble radius. This is an
interesting result, because in this way we might be able to
establish the quantum theory of the spinor perturbations, just as
what is done in inflation theory.

Note that, the above derivative does not mean that Spinor Quintom
is able to avoid any instabilities. We still do not study the
effects of the right hand side of Eq. (\ref{perteq}) which may
destroy the system under some certain occasions. Another possible
instability may be from the quantum effect that our model is
unable to be renormalized. The more detailed calculation will be
investigated in the future works.

\section{A unified model of Quintom and Chaplygin gas}

In the above analysis, we have learned that a spinor field with a
power-law-like potential behaves like a perfect fluid with a
constant EoS. However, it is still obscure to establish a concrete
model to explain how a universe dominated by matter evolves to the
current stage that is dominated by dark energy. In recent years,
another interesting perfect fluid with an exotic EoS $p=-c/\rho$
has been applied into cosmology \cite{Kamenshchik:2001cp} in the
aim of unifying a matter dominated phase where $\rho \propto
1/a^3$ and a de-Sitter phase where $p=-\rho$ which describes the
transition from a universe filled with dust-like matter to an
exponentially expanding universe. This so-called Chaplygin gas
\cite{Kamenshchik:2001cp} and its generalization
\cite{GonzalezDiaz:2002hr} has been intensively studied in a
literature. Some possibilities for this model motivated by field
theory are investigated in \cite{Bilic:2002vm}. A model of
Chaplygin gas can be viewed as an effective fluid associated with
D-branes \cite{Bordemann:1993ep, Fabris:2001tm}, and also can be
obtained from the Born-Infeld action \cite{Bento:2002ps,
Bento:2003we}. The combination of Quintom and Chaplygin gas has
been realized by the interacting Chaplygin gas model
\cite{Zhang:2005jj} as well as in the framework of Randall-Sundrum
braneworld \cite{GarciaCompean:2007vh}.

The Chaplygin gas model has been thoroughly investigated for its
impact on the cosmic expansion history. A considerable range of
models was found to be consistent with SN Ia data
\cite{Makler:2002jv}, the CMBR \cite{Bento:2002yx}, the gamma-ray
bursts \cite{Bertolami:2005aa}, the X-ray gas mass fraction of
clusters \cite{Cunha:2003vg}, the large scale structure
\cite{Bilic:2001cg}, and so on.

Here, we propose a new model constructed by Spinor Quintom which
combines the property of a Chaplygin gas. The generic expression
of the potential is given by
\begin{equation}
V=\sqrt[1+\beta]{f(\bar\psi\psi)+c}~,
\end{equation}
where $f(\bar\psi\psi)$ is an arbitrary function of
$\bar\psi\psi$. Altering the form of $f(\bar\psi\psi)$, one can
realize both the Chaplygin gas and Quintom scenario in a spinor
field.

Firstly, let us see how this model recovers a picture of
generalized Chaplygin gas. We take
$f(\bar\psi\psi)=V_0(\bar\psi\psi)^{1+\beta}$, and then the
potential is given by
\begin{eqnarray}
V=\sqrt[1+\beta]{V_0(\bar\psi\psi)^{1+\beta}+c}~.
\end{eqnarray}
Due to this, we obtain its energy density and pressure
\begin{eqnarray}
\rho_{\psi}&=&\sqrt[1+\beta]{V_0(\bar\psi\psi)^{1+\beta}+c}~,\\
p_{\psi}&=&-c
[V_0(\bar\psi\psi)^{1+\beta}+c]^{-\frac{\beta}{1+\beta}}~.
\end{eqnarray}
Now it behaves like a generalized Chaplygin fluid which satisfies
the exotic relation $p_{\psi}=-\frac{c}{\rho_{\psi}^{\beta}}$. To
be more explicitly, we take $\beta=1$, then get the expressions of
energy density and pressure to be
\begin{eqnarray}
\rho_{\psi}=\sqrt{\frac{N^2}{a^6}+c}~,~~p_{\psi}=-\frac{c}{\rho_{\psi}}~.
\end{eqnarray}
In this case a perfect fluid of Chaplygin gas is given by a spinor
field. Based on the above analysis, we may conclude that this
simple and elegant model is able to mimic different behaviors of a
perfect fluid and so that accommodate a large variety of
evolutions phenomenologically.

In succession, we will use this model to realize a combination of
a Chaplygin gas and a Quintom-A model which is mildly favored by
observations. Choosing $f(\bar\psi\psi)$ to be
$f(\bar\psi\psi)=V_{0}(\bar\psi\psi-b)^{2}$, we get the potential
\begin{equation}\label{case4}
V=\sqrt{V_{0}(\bar\psi\psi-b)^{2}+c}~,
\end{equation}
where $V_{0}$, b, c are undetermined parameters. So we obtain the
derivative of the potential
\begin{eqnarray}
V'=\frac{V_0(\bar\psi\psi-b)}{\sqrt{V_{0}(\bar\psi\psi-b)^{2}+c}}~,
\end{eqnarray}
and the EoS
\begin{eqnarray}\label{eosch}
w_{\psi}=-1+\frac{V_0\bar\psi\psi(\bar\psi\psi-b)}{V_0(\bar\psi\psi-b)^2+c}~,
\end{eqnarray}
respectively, and the crossing over $-1$ takes place when
$\bar\psi\psi=b$.

\begin{figure}[htbp]
\begin{center}
\includegraphics[scale=1.0]{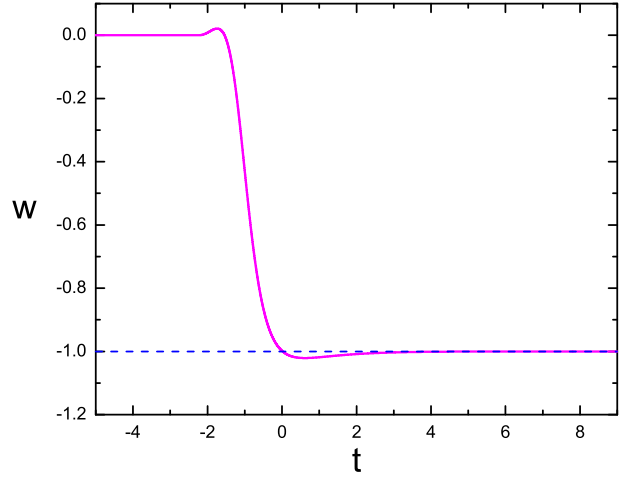}
\caption{Plot of the EoS of the unified model in Eq. (\ref{case4})
as a function of time. In the numerical calculation we take
$V_0=3.0909\times10^{-239}$, $b=0.05$ and $c=9\times10^{-241}$.
For the initial conditions we take $N=0.051$.
\label{Fig:quindeos}}
\end{center}
\end{figure}

\begin{figure}[htbp]
\begin{center}
\includegraphics[scale=0.9]{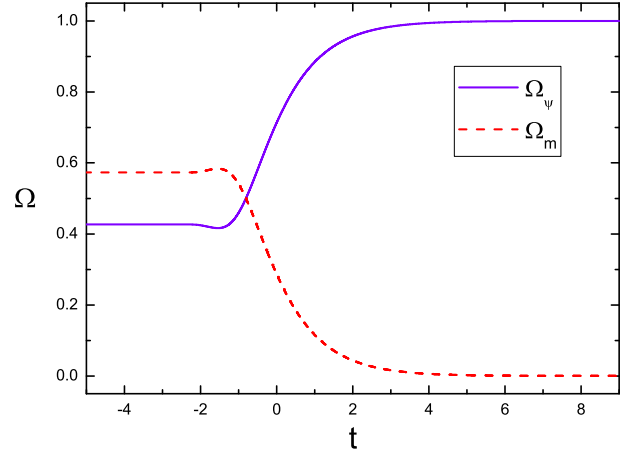}
\caption{Plot of the fraction of energy density of dark energy
(the violet solid line) and normal matter (the red dash line) as a
function of time. \label{Fig:quindomg}}
\end{center}
\end{figure}

During the expansion $\bar\psi\psi$ is decreasing following Eq.
(\ref{solution}). From the formula of the EoS (\ref{eosch}), we
deduce that at the beginning of the evolution the scale factor
$a\rightarrow 0$ so $\bar\psi\psi\rightarrow \infty$. To neglect
the terms of lower order of $\bar\psi\psi$ and the constants, the
EoS at early times evolves from $0$ which describes the era of
matter dominated. Along with the evolution, $w_{\psi}$ increases
from $0$ to the maximum and then starts to decrease. When
$\bar\psi\psi=b$ the EoS arrives at the cosmological constant
boundary $w_{\psi}=-1$ and then crosses it. Due to the existence
of $c$ term, the EoS approaches the cosmological constant boundary
eventually. In this case the universe finally becomes a de-Sitter
space-time. Considering a universe filled with normal matter and
such a Spinor Quintom matter, we take the numerical calculation
and show the evolution of the EoS in Fig. \ref{Fig:quindeos}.
Moreover, we display the evolvement of fraction densities of the
normal matter $\Omega_m \equiv \rho_m/(\rho_m+\rho_{\psi})$ and
Spinor Quintom $\Omega_{\psi} \equiv
\rho_{\psi}/(\rho_m+\rho_{\psi})$ in Fig. \ref{Fig:quindomg}. It
is evident that there is an exact ratio of these two components
from the beginning of evolution, in relief of fine-tuning problems
and accounting for the coincidence problem. Then they evolve to be
equal to each other in late times. Along with the expansion of
$a$, the dark energy density overtakes the matter energy density
driving the universe into an accelerating expansion at present and
eventually dominates the universe completely, which describes an
asymptotic de-Sitter spacetime.

\section{Conclusion and Discussions}

The current cosmological observations indicate the possibility
that the acceleration of the universe is driven by dark energy
with EoS across $-1$, which will challenge the theoretical model
building of the dark energy if confirmed further in the future. In
this paper we have studied various Quintom scenarios in virtue of
a spinor field and proposed a unified model of Spinor Quintom and
a generalized Chaplygin gas. As shown in the present work, this
model can give rise to the EoS crossing the cosmological constant
boundary during the evolution by varying the sign of the term
$V'$. Compared with other models with $w$ across $-1$ in the
literature, so far the present one is also economical in the sense
that it merely involves a single spinor field.

\section*{Acknowledgments}

It is a pleasure to thank Yun-Song Piao, Taotao Qiu, Jun-Qing Xia,
Shiping Yang, Xin Zhang, and Xinmin Zhang for helpful discussions.
CYF acknowledge two anonymous referees for their valuable
suggestions. CYF also thank Haiying Xia and Yi Wang for checking
spelling typos. This work is supported in part by the National
Natural Science Foundation of China under Grants Nos. 90303004,
10533010, 10675136 and 10775180 and by the Chinese Academy of
Science under Grant No. KJCX3-SYW-N2.

\end{document}